\documentclass{sigchi}

\usepackage{etoolbox}
\makeatletter
\patchcmd{\maketitle}{\@copyrightspace}{}{}{}
\makeatother

\usepackage{balance}       
\usepackage{graphics}      
\usepackage[T1]{fontenc}   
\usepackage{txfonts}
\usepackage{mathptmx}
\usepackage[pdflang={en-US},pdftex]{hyperref}
\usepackage{color}
\usepackage{booktabs}
\usepackage{textcomp}

\usepackage{microtype}        
\usepackage{ccicons}          

\usepackage{xcolor}
\usepackage{enumitem}

\definecolor{colandrea}{rgb}{0.0, 0.5, 0.5}

\usepackage{todonotes}

\def\plaintitle{`A Modern Up-To-Date Laptop' - Vagueness in Natural Language Queries for Product Search}
\def\plainauthor{Andrea Papenmeier, Alfred Sliwa, Dagmar Kern, Daniel Hienert, Ahmet Aker, Norbert Fuhr}
\def\plainkeywords{Information retrieval; information need; query formulation; vagueness; natural language.}

\makeatletter
\def\url@leostyle{%
  \@ifundefined{selectfont}{
    \def\UrlFont{\sf}
  }{
    \def\UrlFont{\small\bf\ttfamily}
  }}
\makeatother
\def\pprw{8.5in}
\def\pprh{11in}

\definecolor{linkColor}{RGB}{6,125,233}
\hypersetup{%
  pdftitle={\plaintitle},
  pdfauthor={\plainauthor},
  pdfkeywords={\plainkeywords},
  pdfdisplaydoctitle=true, 
  bookmarksnumbered,
  pdfstartview={FitH},
  colorlinks,
  citecolor=black,
  filecolor=black,
  linkcolor=black,
  urlcolor=linkColor,
  breaklinks=true,
  hypertexnames=false
}

\begin{document}

\title{\plaintitle}

\numberofauthors{2}
\author{%
Andrea Papenmeier\textsuperscript{1}, Alfred Sliwa\textsuperscript{2}, Dagmar Kern\textsuperscript{1}, Daniel Hienert\textsuperscript{1}, Ahmet Aker\textsuperscript{2}, Norbert Fuhr\textsuperscript{2}
  \alignauthor{\vspace{1pt}
    \affaddr{\textsuperscript{1}GESIS – Leibniz Institute for the Social Sciences}\\
    \affaddr{Cologne, Germany}\\
    \email{firstname.lastname@gesis.org}}\\
  \alignauthor{\vspace{1pt}
    \affaddr{\textsuperscript{2}University of Duisburg-Essen}\\
    \affaddr{Duisburg, Germany}\\
    \email{firstname.lastname@uni-due.de}}\\
}

\maketitle

\begin{abstract}
  With the rise of voice assistants and an increase in mobile search usage, natural language has become an important query language. So far, most of the current systems are not able to process these queries because of the vagueness and ambiguity in natural language. Users have adapted their query formulation to what they think the search engine is capable of, which adds to their cognitive burden. With our research, we contribute to the design of interactive search systems by investigating the genuine information need in a product search scenario. In a crowd-sourcing experiment, we collected 132 information needs in natural language. We examine the vagueness of the formulations and their match to retailer-generated content and user-generated product reviews. Our findings reveal high variance on the level of vagueness and the potential of user reviews as a source for supporting users with rather vague search intents.
\end{abstract}

\ccsdesc[500]{Human-centered computing~User interface programming}
\ccsdesc[500]{Human-centered computing~Text input}
\ccsdesc[300]{Human-centered computing~Empirical studies in HCI}

\keywords{\plainkeywords}

\printccsdesc

%
%
%
%
%
%
%

\section{Introduction}

In 2018, \emph{GlobalWebIndex} reported that 27\% of mobile internet users made use of the voice search functionality of their mobile devices (not including tablets) \cite{mander2018voice}, rising from 18\% in 2015 \cite{GWIQ42015}. With the rising numbers of mobile searches that is accompanied by an increase in voice interactions \cite{guy2016searching}, natural language support for searching the Web has gained importance. Likewise, voice assistants such as \emph{Siri}, \emph{Alexa}, or \emph{Cortana} are dependent on understanding and interacting with natural language. Applications of voice assistants in laboratory assistants \cite{cambre2019vitro} or Smart Homes \cite{cho2019once,sciuto2018hey} show the usefulness of voice as an input modality, but, at the same time, highlight existing problems: Conversation techniques are not yet sophisticated enough to elicit long-term usage \cite{cho2019once} and natural language has a great variance in vocabulary \cite{cambre2019vitro}. Already in 1987, Furnas et al. \cite{furnas1987} noted the ``vocabulary problem'': The natural language of users is not equal to the controlled language used to index information in search systems. Although users are best at expressing their \emph{information need} (i.e. what they are searching for) through natural language \cite{lewis1996natural}, traditional information retrieval systems for searching the Web were not designed to deal with challenges such as vagueness and ambiguity \cite{balfe2004improving}. This leads to serious consequences for the users. They need to focus not only on accessing and verbalising their information need, but also on respecting the formal restrictions of the system in order to achieve an acceptable search outcome \cite{ter1996query}. First, this means an increased cognitive burden for the user. Second, the users adapts the formulation of their information need to what they believe the system can process \cite{barsky2005search,kammerer2012children}. This impedes an intuitive interaction with the search system and leads to less relevant search results.

In our research, we contribute to the vocabulary problem in web search by designing an interactive search system from a user-centric view that is able to handle users’ information needs with all the vagueness and ambiguity that might be included. In this paper, we report on our first step in a user-centred design process: the collection and investigation of natural language descriptions of information needs. We designed and conducted a user study with 132 participants to collect genuine information need descriptions in the product search context. We examined the vagueness of these natural language descriptions and explored how well they match with retailer-generated content (product descriptions) as well as user-generated content (user reviews). Our key findings show that 
user reviews have a high potential as a source for matching vague descriptions of products. User reviews have syntactic and semantic similarities to the vague information needs and contain information about users’ experiences with a product that are not provided in retailer-generated descriptions (e.g. quality or brand reputation). We conclude our paper by discussing implications for designing more intuitive product search systems. 

\section{Related Work}
In the following section, we present the current state of research on natural language in voice interaction and for interacting with search systems, as well as existing approaches to resolve challenges arising from natural language for querying. 

\subsection{Natural Language in Voice Interactions}
The quality of processing natural language is an essential driver for high user experience in area of voice assistants. Voice assistants have been evaluated in many scenarios \cite{cambre2019vitro,cho2019once,robb2019exploring,vtyurina2018exploring}, highlighting the potential for conversational interaction, but also showing existing challenges. Cambre et al. \cite{cambre2019vitro} employed a voice assistant in a laboratory setting, noting the challenge of the versatile natural language vocabulary. In their use case, the system was unable to understand technical terms used during laboratory work. Missing context is also a problem often described in literature \cite{cambre2019vitro,klopfenstein2017rise,robb2019exploring}. To date, interacting with voice assistants is mostly restricted to simple commands and implemented ``skills''. Yet, in an ideal setting, voice interaction is highly versatile and conversational \cite{vtyurina2018exploring}. Interaction with voice assistants is therefore not naturally restricted to simple commands or a closed vocabulary. The absence of sophisticated conversational abilities and lack of understanding the rich natural language often results in disappointment and frustration, not only for voice assistants but also conversational chat bots \cite{jain2018evaluating}. In general, voice results in a strong anthropomorphisation \cite{cho2019once}. Cho, Lee and Lee \cite{cho2019once} conducted a long-term study to investigate why long-term usage of \emph{Amazon's} voice assistant \emph{Alexa} is low. Participants unconsciously anthropomorphised the virtual assistant, leading to disappointment when \emph{Alexa} did not live up to the expectation of a human-human-like conversation. In human-human interaction, vagueness does not necessarily lead to a problem of understanding. Jucker, Smith and L{\"u}dge \cite{jucker2003interactive} argue that vagueness is a result of using an appropriate level of detail with respect to a specific recipient and situation. It can therefore be considered a ``tool'' to reduce cognitive burden.

\subsection{Query Formulation Problem}
Traditional search engines require users to issue a query at the beginning of the search process. Previous research argued that users modify and reformulate the query if they are not satisfied with the results \cite{ter1996query}. Kato et al. \cite{kato2014cognitive} analysed logs from a search engine and found that experienced searchers adapt their formulation to what they believe the search system is able to process. In the same vein, Kammerer and Bohnacker \cite{kammerer2012children} analysed the query formulation process of children (age 8-10 years). In their experiment, younger and inexperienced children tended to use natural language rather than keyword search, while older children had already learned to use keywords as their search strategy. Aula \cite{aula2003query} likewise found familiarity with the search system and domain experience to be influencing query formulations. For search systems, vagueness is a major problem: Balfe and Smyth \cite{balfe2004improving} argue that due to their briefness, search queries lack the necessary context to restore the user's information need. Introducing controlled languages (e.g. boolean logic) to overcome the query formulation problem did not show promising results in the past, as they are often applied incorrectly \cite{jansen1998failure}.

In the context of voice search, relatively little research has been conducted on query formulation. Guy \cite{guy2016searching} analysed the query logs of a voice search system by comparing them to textual queries, revealing some subtle differences between the two input modalities. Voice queries contain more often words that are easy to pronounce but difficult to type. Yet, the opposite could be observed with typed queries, which contained more often words that are easy to type but lengthy to verbalise, e.g. calendar years. Guy \cite{guy2016searching} also found that language in voice queries is richer, i.e. more varied, and closer to natural language than typed queries.

The evidence presented so far suggests that information need can best be expressed in natural language, as opposed to controlled query languages. However, users are influenced in their query formulation process by their beliefs about what a search system is capable of. Since information retrieval systems traditionally have not been designed to support vagueness arising from natural language, examining vagueness in queries will not be possible via logs of existing search engines.

\subsection{Faceted Search}
Several researchers have investigated methods to support the user in searching large amounts of data, e.g. in product search. Faceted search provides filtering opportunities while exploiting additional, structured information about the items \cite{hearst2002finding,yee2003faceted}. These facets are generated based on provided product items. Approaches are suggested to better adjust these facets to user needs, e.g. by changing the ordering of facets \cite{koren2008personalized} or by automatically extracting relevant facets with respect to the user query \cite{tunkelang2006dynamic}. Adding a ``weight'' to a facet can further help the users to indicate how much impact a facet should have on the result list \cite{kern2018evaluation}. Research has also shown that novice users profit from different content than experts: In e-commerce, novices need more general information about a product than experts, who prefer more detailed information \cite{sawasdichai2002user}.

Recent research has focused on data-driven improvements of product search. Hirschmeier and Egger \cite{hirschmeier2018social} built facets from notebook reviews on \emph{Amazon} and noted that user-generated content holds \textit{``a problem and an opportunity at the same time''} \cite{hirschmeier2018social}: For a user, there is too much content to consider in detail. At the same time, it provides enough data to automatically extract product attributes that are frequently mentioned. However, a user study to validate the usability of their extracted facets is missing. Similarly, Feuerbach et al. \cite{feuerbach2017enhancing} used reviews to generate new facet values for existing facets of a hotel search system, e.g. \textit{``comfortable''} for the facet \textit{``bed''}. In a user study with 30 participants, they found that users perceived the extracted facet values more suitable with respect to their preferences as compared to the original values. In their work on ontologies for multilingual product search, Lehtola, Heinecke and Bounsaythip \cite{lehtola2003intelligent} investigated extraction and mapping of colour and material attributes for clothes. They highlight the discrepancy between retailers (synonyms are used for the same material), but also between retailer and customer (preference for the name of a textile vs. its functionality).

Kleemann and Ziegler \cite{kleemann2019integration} developed a dialog-based product advisor for notebooks. The advisor poses questions about abstract attributes of the product rather than asking about precise technical characteristics, e.g. \textit{``What do you want to use it for?''} (translated from German) with the answer options \textit{``surfing and watching movies''}, \textit{``gaming''}, \textit{``advanced office tasks''}, and \textit{``image and video editing''}. Vaccaro et al. \cite{vaccaro2018designing} likewise take a user-centered approach for their work on personal fashion assistants. They analysed the interaction between personal fashion advisors and their clients to develop design guidelines for fashion assistant chatbots. These guidelines differ from the abstract questions used by Kleemann and Ziegler \cite{kleemann2019integration}, leading to the assumption that a difference between those two domains, the technical and the clothes domain, exist with regards to product search support. Sawasdichai and Poggenpohl \cite{sawasdichai2002user} explored e-commerce shopping from the perspective of the users and found that users, in general, expect to be provided with the same information when shopping offline as they are in an offline setting.

Although these works show that researchers experiment with different information sources to improve product search, it remains unclear whether the improvements fit the user's language. First user studies (evaluating for example a natural language product advisor or new facet values) show the positive perception by users, but the development process did not start with the users in the first place. Little research has focused on the user's genuine information need as a source for identifying product attributes for faceting and filtering.

\section{Research Method}
In the context of designing an interactive search system following a user centerd design process, we conducted a first user study to investigate the formulation of user's information need in natural language and their potential for improving product search systems. Since users adjust their query to the search system, analysing query logs is not sufficient for investigating natural language as query language. Therefore, we decided to carry out a user study in which the formulation of the information need is decoupled from an existing search system. We focus on two different domains: the technical domain and clothes domain. For the technical domain, we chose purchasing a laptop as an example and for the clothes domain our decision went in favour of jackets.

Our user study is designed and conducted as an online survey to collect real-life examples of information need formulations. The study is set up as an in-between study with two independent conditions (laptop domain, jacket domain). The participants are asked to describe a product (independent of the search context), before retrieving the described item from an online product search website.

Our research is driven by the following research questions:

{\small \textbf{RQ 1}} How do users formulate their information need in natural language in a product search scenario?
\begin{enumerate}
    \item [] {\small \textbf{RQ 1.1}} How vague do users formulate their query in product search when using natural language?
    \item [] {\small \textbf{RQ 1.2}} What role does the product domain play for query formulation?
\end{enumerate}
{\small \textbf{RQ 2}} To what extent do retailer-generated contents and product reviews reflect the language used in user-generated product descriptions?

{\small \textbf{RQ 3}} How well do facets of online shops match the user-generated product description?

\section{Study Design}
The following section describes the design and sample for the online user study.

\subsection{Scenario and Task}
We re-use the task described by Barbu el al. \cite{barbu2019influence} who investigated the impact of review tonality on buying decisions within product search. In their study, participants are asked to imagine themselves ``looking to purchase a new laptop after their old one broke''. Except the target product, this task does not prime the participants to re-use formulations of the task description. We adapt the scenario for the technical domain:
\begin{quote}
    \textit{Imagine your laptop broke down. What kind of laptop would you choose as a replacement? Please describe the laptop you would want as a replacement in your own words.}
\end{quote}
For the clothes domain, we alter the scenario to match the need for a new jacket, while keeping the instructions constant:
\begin{quote}
    \textit{Imagine you lost the jacket you wear on a daily basis. What kind of jacket would you choose as a replacement? Please describe the jacket you would want as a replacement in your own words.}
\end{quote}
In both cases, participants are instructed to write at least 50 characters, serving two goals. First, texts with a certain length can be used as an attention check (i.e. it shows whether the participant has understood the instructions correctly), and secondly, we expect that requesting longer texts will elicit a natural language rather than a bullet points.

After formulating their information need, participants were asked to search for the product on \emph{Amazon}. \emph{Amazon} was chosen as a well-known representative of a non-specialised online product retailer. To avoid superficial searches, participants are instructed to search for approximately five minutes. This time frame is weakly enforced by the survey system by blocking the ``next'' button for two minutes. At the end of the search, participants report the URL to their final choice. To investigate how much the search has influenced the users' search intent, we offer participants the opportunity to make changes to their initial product description.

\subsection{Apparatus \& Procedure}
To reach native English speaking participants, it is set up as an online study. We use the survey platform SoSci Survey\footnote{https://www.soscisurvey.de}. For both the survey and the search task, participants are asked to use the internet browser on their private device. The study is structured as follows:
\begin{enumerate}
    \item Introduction and consent
    \item Scenario and task: multi-line text field for information need description
    \item Product search task: link to \emph{Amazon}, and URL input field
    \item Description refinement: display of participant's answer of step 2, with possibility to refine and change the description
    \item Post-task questionnaire: query, domain knowledge, and last product search in the specific domain
    \item Demographic questions: gender, age
\end{enumerate}
The descriptions are manually evaluated by the authors to verify that the task has been read attentively and understood correctly.

\subsection{Measures}
\label{subsec:measures}
Per participant, we measure the following dependent variables: 
(1) information need description, (2) search query, (3) domain knowledge, (4) the search outcome in the form of the URL to the product on \emph{Amazon}, (5) satisfaction with the search outcome, and (6) search duration. The information need description, search query, and product URL are provided by the participant in full-text, while the search duration is automatically recorded by the survey system. We measure domain knowledge as a combination of search experience and self-assessment questions, as suggested by Kanwar, Grund and Olson \cite{kanwar1990measures}. We ask the participant to indicate how much they think they know about several product attributes (as found in common facets of online shops) on a 7-point polarity scale (with 1 = ``no knowledge'', 7 = ``expert knowledge''). 
Satisfaction with the search outcome is measured on a 7-point Likert scale (from -3 = ``very dissatisfied'' to +3 = ``very satisfied''). In a second step after the user study, user-generated descriptions are annotated with a ``vagueness score''. To investigate how well existing systems support natural language queries, we collect retailer-generated product descriptions and user-generated product reviews (\textbf{RQ 2}) as well as the facets of online product retailers (\textbf{RQ 3}). For each of these information sources, we calculate how well they match the user-generated descriptions.

\subsection{Statistical Tests}
For all significance tests between two independent samples (e.g. laptop domain vs. jacket domain or user-generated content vs. retailer-generated content), we use the two-sided Mann-Whitney U-test.  
The correlation of dependent variables is computed using Pearson's correlation coefficient, while the significance of differences between two paired samples is evaluated with Wilcoxon's signed-rank test.

\subsection{Participants}
In total, 149 participants were recruited on the scientific crowd sourcing platform Prolific\footnote{https://www.prolific.co}. 17 participants had to be excluded from the analysis due to misunderstanding the task. The final sample size is therefore N = 132 (f = 83, m = 49, d = 0). Participants had to be older than 18 years (M = 34.1 years, SD = 11.2 years) to take part in the experiment and each received a financial allowance of 0.80 GBP for 7 minutes of work (6.80 GBP per hour). The platform's population was screened for participants with English as their first language to avoid a translation bias.  
Furthermore, participants were informed before the start that they will need to access \emph{Amazon.com} in the course of the experiment to avoid technical difficulties. Participants are equally distributed over the two domains with respect to age and gender, with 66 participants searching for a laptop and 66 searching for a jacket.

\section{Data Preparation}
This section describes the text preprocessing, segmentation and annotation process of the user-generated product descriptions, as well as the collection of lists of existing facets from a general retailer and specialised retailers for both domains (laptop and jacket).

\subsection{Annotations}
As described in the \hyperref[subsec:measures]{Measures section}, we enrich the user-generated information need descriptions with a vagueness score. We followed the method used by Lebanoff and Liu \cite{lebanoff2018automatic} who use crowd sourcing with native English speakers to annotate the level of vagueness of a sentence. Four annotators were recruited on Amazon Mechanical Turk\footnote{https://www.mturk.com} to each label all user-generate descriptions on a scale from 1 (``very specific, not vague at all'') to 10 (``very vague, not specific at all''). They received a compensation of 7.00 USD for one hour of work and had to have an approval rate on Mechanical Turk of more than 95\%. Before labelling the dataset, annotators were introduced to our definition of vagueness:
\begin{quote}
    \textit{Vagueness is the imprecise or unclear use of language. Contrast this term with ``clarity'' and ``specificity''. Vague language states a general idea but leave the precise meaning to the reader's interpretation.}
\end{quote}
The annotators performed a training phase, in which they annotated nine descriptions from pretests to get used to the topic, the annotation scale, and the formulations. With an estimated duration of an hour, we anticipated the task to be rather lengthy. To ensure high-quality annotations, we included three attention check mechanisms (minimum duration of 20 minutes, asking the participant after the training phase for the maximum scale value, and a question in between the annotations asking to tick a specific value). The attention checks disqualified one annotator.

The three annotators had a reliability measured with Krippendorff's $\alpha$ of .62. The average of their individual test-retest reliability is at Krippendorff's $\alpha$ = .67, while the Pearson's correlation coefficient among the three annotators is at r(1) = .63 with p < .001 in all cases. Although the inter-annotator reliability is not very high, the annotators show a high correlation and an acceptable test-retest reliability. We expected judging the vagueness on a sentence level to be an ambiguous task, which reflects in rather low reliability scores. The annotator reliability is comparable if not better to the one in \cite{lebanoff2018automatic}, who report that 4/5 of their annotators agreed on 13\% of the descriptions, which, in our case, is 13\% for 3/3 annotators. In 47\% of their cases, 3/5 annotator agreed, while in our case 2/3 annotators agreed on 70\% of the cases. Note that for this comparison, we mapped our 10-point scale to their 5-point scale. As done in \cite{lebanoff2018automatic}, we average the scores of all three annotators to obtain a single decimal number as vagueness score for each description.

\subsection{Segmentation}
To see the influence of the search task on the information need, participants had the opportunity to modify their description after the search process. However, for all analyses besides the direct comparison of initial and modified description, we draw on the initial, uninfluenced descriptions. For investigating the descriptions with respect to product attributes, the initial information need descriptions are split into segments. Each segment (line) contains information on exactly one product attribute (in bold), e.g.:
\begin{quote}
    \textit{One that's \textbf{lightweight},\\
    \textbf{warm}\\
    and has a \textbf{hood}.}
\end{quote}
Independent of each other, two of the authors and a second person from outside the research project divided the user-generated descriptions into segments. The final segmentation was determined via majority vote (90\% of the descriptions). If all three segmentations of a description differed from each other (10\% of the descriptions), the group discussed the segmentation until a consensus was found. Three rules were set before the segmentation process: 1) no deletion of words, only delimitation of the text, 2) except for ``and'', ``but'', and ``or'' if used as conjunction between two attributes, 3) duplication of words that refer to two attributes, e.g. \textit{``fast start up and use''} to \textit{``fast start up''} and \textit{``fast use''}. Overall, 132 descriptions were segmented into 570 segments (252 in laptop descriptions, 318 in jacket descriptions), each containing one product attribute. The descriptions contain at least one attribute and at most 10, with 4 attributes on average (M = 3.82, SD = 1.82).


\subsection{Text Preprocessing}
Before comparing the information need descriptions to retailer-generated content as well as user reviews, we preprocess each text. For the information need descriptions, we manually sort out text fragments that do not contain information on the desired product characteristics but are artifacts of natural language sentence structure and the task instructions, e.g. \textit{``I would like''} or \textit{``I would try to find''} or stop words such as \textit{``a''} or \textit{``and''}. This step was taken to avoid false positives when unimportant words are matched, and false negatives due to unsubstantial words lacking in the target text. The automatic preprocessing steps for all texts include: (1) conversion of texts to lowercase, (2) removal of trailing characters, (3) removal of punctuation characters, (4) lemmatization of each word in the texts and (5) removal of stop words. The preprocessing pipeline is realised using the NLTK\footnote{\url{https://www.nltk.org/api/nltk.html}} library.

\subsection{Facet Matching}
To answer \textbf{RQ3} ``How well do facets of online shops match the user-generated product description?'', we collected lists of product search facets. We retrieve the facets for each domain from both a general retailer (\emph{Amazon}) and a specialised retailer (\emph{skinflint} for laptops and \emph{next} for jackets). To avoid influence on the facet lists by search queries, we navigate to the product categories via the websites' menu bars. On the website of the general retailer \emph{Amazon}, we navigate to the domain categories in a private browser (``Shop by category'' > ``Computers'' > ``Computers \& Tablets''  > ``Laptops'' and ``Shop by category'' > ``Men's Fashion'' > ``Clothing''  > ``Jackets \& Coats'', same path for ``Women's Fashion'') and retrieve the list of facets with facet values. On \emph{skinflint}, the specialised retailer for technical products, we navigate to the laptop category via ``Hardware'' > ``Notebooks''  > ``Notebooks''. For retrieving specialised jacket facets on  \emph{next}, we follow ``Women'' > ``Clothing''  > ``Coats \& Jackets'' (or starting at ``Men'', respectively). The gender-specific facets for jackets are merged into a single list of facets. The final lists of the general retailer contain 27 facets for laptops and 12 for jackets. The specialised retailer lists offer 84 facets for laptops and 14 facets for jackets. Table \ref{tab:facets} shows the first five facets on each list, sorted alphabetically, while the full lists are available online\footnote{\url{https://git.gesis.org/papenmaa/dis20_usersearchintentformulation}}.

\begin{table}[tb]
	\centering
	\begin{tabular}{p{0.1\textwidth}p{0.1\textwidth}|p{0.09\textwidth}p{0.1\textwidth}}
	    \multicolumn{2}{c|}{\textbf{Laptop}} & \multicolumn{2}{c}{\textbf{Jacket}} \\
	    \textbf{Amazon} & \textbf{skinflint} & \textbf{Amazon} & \textbf{next} \\ \midrule
	    
	    \begin{itemize}[leftmargin=*, noitemsep]
	    \renewcommand\labelitemi{-}
	        \item Aspect ratio
	        \item Battery
	        \item Card Readers
	        \item Class
	        \item Code-name AMD 
	    \end{itemize}
	    & 
	    \begin{itemize}[leftmargin=*, noitemsep]
	    \renewcommand\labelitemi{-}
	        \item Activity
	        \item Average Customer Review
	        \item Certifi-cations
	        \item Condition
	        \item CPU Speed
	    \end{itemize}
	    &
	    \begin{itemize}[leftmargin=*, noitemsep]
	    \renewcommand\labelitemi{-}
	        \item Benefit
	        \item Brand
	        \item Category
	        \item Colour
	        \item Design Feature
	    \end{itemize}
	    &
	    \begin{itemize}[leftmargin=*, noitemsep]
	    \renewcommand\labelitemi{-}
	        \item Big \& Tall Size
	        \item Brand
	        \item Color
	        \item New Arrivals
	        \item Petite Size
	    \end{itemize}
	    \\ \bottomrule
	\end{tabular}
	\caption{First five entries of facet lists for both domains and both the unspecialised (\emph{Amazon}) and specialised (\emph{skinflint}, \emph{next}) retailers, ordered alphabetically.}
	\label{tab:facets}
\end{table}

\subsection{Product Page Matching}
During the user study, participants delivered an \emph{Amazon} URL of their chosen product, which is used to crawl the retailer-generated content for the product (i.e. product title and product description) immediately after the study. For each product on \emph{Amazon}, there exist specific HTML-fields for the product title and product description. If available, we also crawled the product's review texts. As products might have multiple reviews, we concatenate the associated reviews to one full review. Eventually, we gather three information sources to describe a product: the title, product description and reviews. These are used for comparison with the user-generated descriptions from the user study for matching purposes. We leverage the text preprocessing pipeline on the three corpora as proposed in the Section ``Text Preprocessing''.

Table \ref{tab:dataset_statistics} illustrates the vocabulary sizes for each information field in the respective domain. This table shows that the vocabulary of retailer-generated content (both product title and product description) in the laptop domain is greater than that in the jacket domain. Contrarily, the vocabulary of reviews in the jacket domain is greater than in the laptop domain.

In the matching step, we determine for each attribute in every description whether the attribute can be found in the different information fields (product title, product description, user reviews). The matching is binary per attribute and determined by simple substring matching. We evaluate the quality of the automatic matching by manually matching and achieve a strong correlation (Pearson's correlation coefficient of r(130) = .63), meaning that the automatic matches can be used to approximate the manual matches. In case of the manual matching, a more extensive and semantically focused matching was done as compared to the strict automatic substring matching, e.g. \textit{``has four pockets''} was matched for the attribute \textit{``multiple pockets''} in the manual matching, although missing the word \textit{``multiple''}. For better reproducibility, we report results based on the automatic matches in the remainder of the paper.

Finally, we compute the ``coverage'' of each user-generated description.  The coverage indicates how many attributes of a description are found in an information field and is calculated as follows: 
\[
\frac{\#~ of~ description~ attributes~ found~ in~ the~ information~ field}{total ~\#~ of~ description~ attributes}
\]
In the description ``\emph{A waterproof and weatherproof jacket in a subtle earthy colour}'', only the attribute ``\emph{colour}'' was found in the list of \emph{Amazon} facets. As the description contains three attributes (``\emph{waterproof}'', ``\emph{weatherproof}'', and ``\emph{subtle earthy colour}''), the coverage of the \emph{Amazon} facets are 33\%.

\begin{table}[tb]
	\centering
	\begin{tabular}{l|cr|crl} 
	    & \multicolumn{2}{c|}{\textbf{Laptop}} & \multicolumn{2}{c}{\textbf{Jacket}} \\ 
	    & \textbf{\# texts} & \textbf{voc. size} &  \textbf{\# texts} & \textbf{voc. size} \\ 
	    \midrule
	    \ titles         & 66 & 296 & 66 & 267\\
	    \ descriptions   & 66 & 1,132 & 66 & 973\\ 
	    \ reviews        & 60 & 14,151 & 55 & 17,532\\ 
	\end{tabular}
	\caption{Statistics about the different corpora regarding to the three information fields for both domains. Vocabulary size refers to the number of unique terms in one corpus.}
	\label{tab:dataset_statistics}
\end{table}


%
%
%
\section{Results}
In the following section, we describe characteristics of the user-generated product descriptions as given before the search, compare the descriptions to the issued queries, and examine how users adjusted the description after the search. In a second step, we present the results of investigating how well the natural language descriptions match to the seller-generated content, to the product reviews given by other buyers, and to the facets currently available in popular product search systems. The complete dataset of the user study, including the segmentations and annotated vagueness scores, is publicly available\footnote{\url{https://git.gesis.org/papenmaa/dis20_usersearchintentformulation}} 
.\newline

\subsection{Vagueness in User Descriptions}
\begin{figure}[tb]
    \centering
    \includegraphics[width=0.45\textwidth]{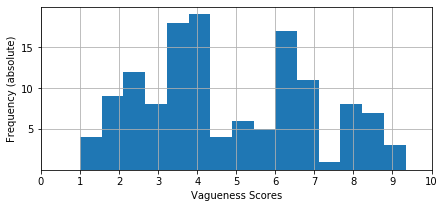}
    \caption{Histogram of vagueness annotations for all 132 user-generated descriptions.}
    \label{fig:vagueness}
\end{figure}

Figure \ref{fig:vagueness} shows the histogram of annotated vagueness of all 132 user-generated product descriptions. The data does not have a normal distribution (Shapiro-Wilk normality test with p = .002), with relatively few descriptions classified at the mean vagueness score (M = 4.79, SD = 2.10). Descriptions are either assigned a rather low vagueness or high vagueness. The domains show a weak significant difference in means (p = .043), with the laptop descriptions being rated more vague (M = 5.10, SD = 2.16) than the jacket descriptions (M = 4.48, SD = 2.00). The most vague laptop description collected in our study (unprocessed) was:
\begin{quote}
    \textit{I'd go for one that is reasonably priced, with a good sized hard drive and ram} (vagueness = 9.3)
\end{quote}
In the jacket domain, the following description was rated the most vague:
\begin{quote}
    \textit{I would want a waterproof jacket that is cosy and warm.} (vagueness = 8.67)
\end{quote}
Whereas the least vague jacket description in our dataset is the following:
\begin{quote}
    \textit{It is a mustard coloured padded jacket, with quite a high collar that has a hood inside it. The cuffs are ribbed at the sleeves. It also has two quite deep side pockets.} (vagueness = 1.0)
\end{quote}
Some user-generated product description were rated to be in the middle of the vagueness scale, e.g.:
\begin{quote}
    \textit{A laptop with a screen over 14 inches and that was light. The brand wouldn't be particularly important but one which looks stylish} (vagueness = 6.0)
\end{quote}

\subsection{User Descriptions and Queries}
On average, the user-generated descriptions of products were 29 words long in the laptop domain, and 24 words long in the jacket domain. The initial search queries used to retrieve the respective product from \emph{Amazon.com} only had a length of 2.2 words for laptops and 2.8 words for jackets, showing that on average, the queries were 92\% (88\%, respectively) shorter than the descriptions. Two participants reported to not have used the search bar and therefore were not able to report a query string. Out of 66 participants in the laptop domain, 21 used the query \textit{``laptop''} or \textit{``laptops''}. On average, 51\% of the query terms also appeared in the description. In the jacket domain, only 3 participants used the queries \textit{``jacket''} or \textit{``coat''} and on average, 46\% of the query terms also appeared in the description they wrote before. In general, queries were much shorter than the user-generated description given before the search. 40\% of participants searching for laptops issued a query with no overlap to their initial description. In the jacket domain, this could be observed in only 21\% of the cases. We observe three types of phenomena when queries contained words not appearing in the description: (1) The usage of \emph{pronouns} instead of nouns, e.g. the query \textit{``laptop''} with the description \textit{``a simple one that does the basics large memory and simple to use''}. (2) \emph{Additional} information in the query, e.g. the query \textit{``navy wool coat''} with the description \textit{``simple and classic navy blue knee lenght coat with a collar made by cos or arket''} where the term \textit{``wool''} was added. (3) \emph{Omission of a word} in the description that is used as a generalisation to summarise the description, e.g. the query \textit{``laptop''} with the description \textit{``14in screen with 1tb memory must include 365 microsoft''}.

After the search task, participants were offered the possibility to adjust their initial product description. 23\% of the participants followed this offering in both domains, whereof 80\% expanded the description, while 20\% shortened it. In those cases, the final description in the laptop domain was 40\% different from the initial description (22\% in the jacket domain), for example by adding \textit{``with a hood''} to \textit{``a light waterproof neutral color jacket brand name but not too expensive''} or changing \textit{``enough ram and graphics card lots of internal harddrive storage''} to \textit{``ram graphics card i7 processor at least 500gb internal storage''}. 

\subsection{Statistical Analysis of Variables}

\begin{table}[tb]
	\centering
	\begin{tabular}{l|rrr|r}
	    & \textbf{All} & \textbf{Laptop} & \textbf{Jacket} & \textbf{Domains} \\ 
	    & \textbf{mean} & \textbf{mean} & \textbf{mean} & \textbf{p-value} \\ \midrule
	    satisfaction    & 5.69 & 6.06 & 5.32 & ** < .001 \\
	    dom. knowl.     & 4.19 & 3.88 & 4.51 & ** .003 \\
	    \# attributes      & 4.33 & 3.83 & 4.83 & ** < .001 \\
	    \# words         & 26 & 29 & 24 & * .012 \\
	    search time     & 140 & 248 & 205 & .372 \\ \bottomrule
	\end{tabular}
	\caption{Means of dependent variables and p-values of test for significance between domains, where a single asterisk denotes significance at 95\% CI and a double asterisk significance at a 99\% CI.}
	\label{tab:results_means}
\end{table}

Table \ref{tab:results_means} reports the means of dependent variables (vagueness, satisfaction, domain knowledge, amount of attributes, amount of words, and search duration). In the last column of Table \ref{tab:results_means}, the results of testing for significant differences within the two domain groups can be found. The statistical test for difference between the laptop and jacket domain indicates that the domain has a significant impact on the query formulation in terms of amount of words, amount of attributes, and satisfaction. Laptop descriptions contain significantly more words ($M_{laptop}$ = 29, $M_{jacket}$ = 24), but significantly fewer attributes ($M_{laptop}$ = 3.83, $M_{jacket}$ = 4.83), which means that users take on average more words to describe a single attribute in the laptop domain. The domain knowledge of participants is reported to be significantly higher in the jacket domain than in the laptop domain ($M_{laptop}$ = 3.88, $M_{jacket}$ = 4.51), while the satisfaction with the result is higher of participants who searched for a laptop as compared to those who searched for a jacket ($M_{laptop}$ = 6.06, $M_{jacket}$ = 5.32). Performing a correlational analysis of the dependent variables, we see that a moderate negative correlation exists between vagueness and amount of attributes mentioned in the description (r(130) = -.444): The more attributes are mentioned, the lower the vagueness of the description. It appears that satisfaction is weakly negatively correlated with the amount of mentioned attributes (r(130) = -.224), meaning that the more attributes the user described before the search, the lower the satisfaction after the search. A very weak positive correlation can be observed between domain knowledge and the amount of mentioned attributes (r(130) = +.187), and a very weak negative correlation between domain knowledge and the search duration (r(130) = -.142).


%
\subsection{Matching Measurements}
\begin{table}[tb]
	\centering
	\begin{tabular}{l|rrr|r}
	    & \textbf{All} & \textbf{Laptop} & \textbf{Jacket} & \textbf{Domains} \\
	    & \textbf{mean} & \textbf{mean} & \textbf{mean} & \textbf{p-value} \\ \midrule
        titles		& 14\% & 15\% & 13\% & .186 \\
        descriptions		& 9\% & 5\% & 14\% & ** < .001 \\
        reviews		& 27\% & 21\% & 33\% & * .011 \\
        titles+descr.		& 18\% & 16\% & 20\% & * .012 \\
        titles+descr.+rev.		& 34\% & 28\% & 41\% & ** .001 \\
	    \bottomrule
	\end{tabular}
	\caption{Average coverages for various information sources, with coverage being the percentage of attributes matched per user-generated description. Right column: p-values of test for significance between domains, where * denotes significance at 95\% CI and ** significance at a 99\% CI.}
    \label{tab:results_coverages_nonfacets}
\end{table}

\begin{figure}[b]
    \centering
    \includegraphics[width=0.47\textwidth]{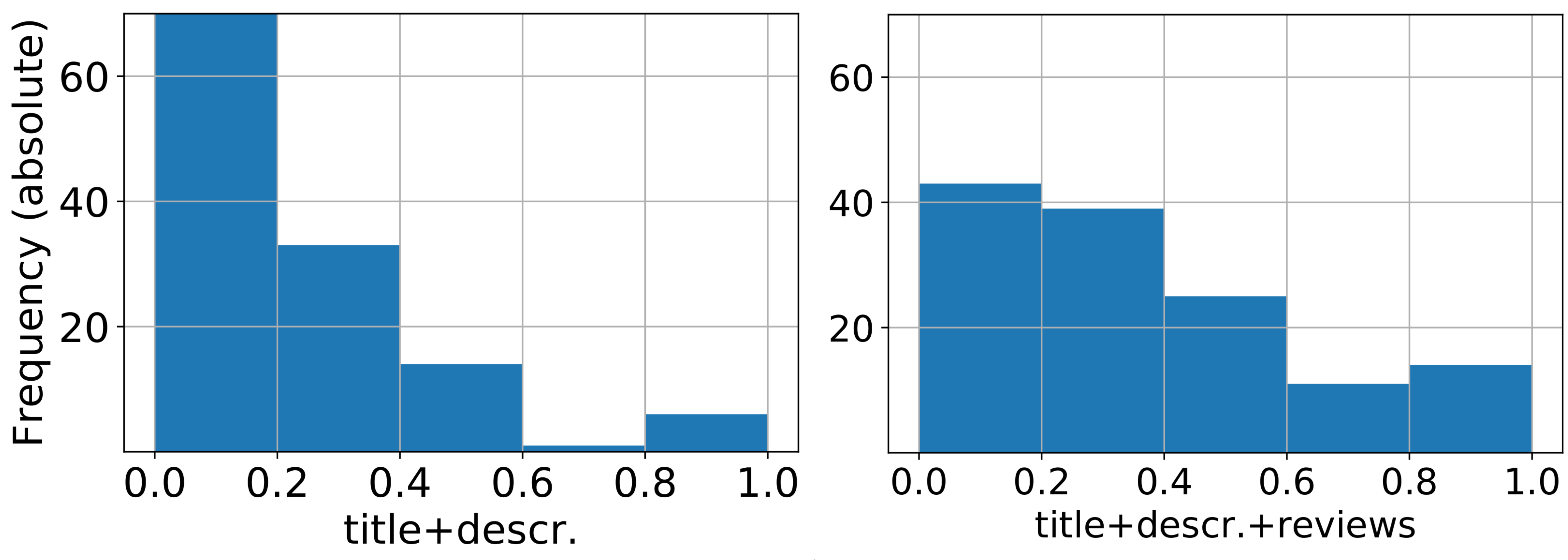}
    \caption{Histograms of matched attributes per user-generated description with respect to product title and product description (left) and with respect to product title, product description, and user reviews (right).}
    \label{fig:matching}
\end{figure}

The average coverage of different information sources is presented in Table \ref{tab:results_coverages_nonfacets} and visualised in Figure \ref{fig:matching} for matching with retailer-generated content (left) and retailer-generated content enriched with user reviews (right). The data in Table \ref{tab:results_coverages_nonfacets} shows a weakly significant higher coverage through retailer-generated content (product page title + product description) in the jacket domain (20\%) as compared to the laptop domain (16\%). Adding user reviews as information source increases this difference: user-generated descriptions are better matched in the jacket domain (41\%) than in the laptop domain (28\%) when considering all information sources (product page title + product descriptions + user reviews).

\subsection{Retailer-Generated Content vs. User-Generated Content}

\begin{table}[tb]
	\centering
	\begin{tabular}{l|rr|r}
	    & \textbf{Low-vag.} & \textbf{High-vag.} & \textbf{} \\
	    & \textbf{mean} & \textbf{mean} & \textbf{p-value} \\ \midrule
        titles		& 31\% & 20\% & ** .003\\
        descriptions		& 42\% & 27\% & ** <.001\\
        reviews		& 44\% & 38\% & .257\\
        titles+descr.        & 23\% & 12\% & ** .007\\
        titles+descr.+rev. & 37\% & 30\% & .088\\
	    \bottomrule
	\end{tabular}
	\caption{Average coverages for various information sources, with coverage being the percentage of attributes matched per user-generated description. Right column: p-values of test for significance between domains, where * denotes significance at 95\% CI and ** significance at a 99\% CI.}
    \label{tab:results_coverages_vagueness}
\end{table}

The data in Table \ref{tab:results_coverages_nonfacets} show that the coverage of retailer-generated content (product page title and product description) can be improved when adding the product reviews: 11 percentage points are added in the laptop domain and 21 percentage points in the jacket domain. The Wilcoxon signed rank test shows significance in both domains with p < .001. The impact of adding reviews for matching becomes apparent in Figure \ref{fig:matching}: The amount of descriptions with a low coverage (between .0 and .2) decreases greatly. Still, even when including user reviews as information source, some individual attributes remain unmatched. Some unmatched attributes are too precise to be mentioned by any information source, e.g. \textit{``that I could wear for dog walking and going out as well as to hockey''} or \textit{``if it came with some extra free programmes like Office I would be overjoyed''}. Other unmatched attributes are highly vague in the formulation, e.g. \textit{``has a bit of longevity''} or \textit{``with a nice big screen''}. 

To determine the benefit of adding reviews for automatically processing vague language, we examine statistical differences between high-vagueness descriptions and low-vagueness descriptions. We divide the user-generated description in two vagueness groups: a low-vagueness group with vagueness scores smaller than the average vagueness score of 4.79 (N=74), and a high-vagueness group with scores greater than the average (N=58). The results are presented in Table \ref{tab:results_coverages_vagueness}. For the low vagueness group, the coverage of retailer-generated content (product page title + product description) is significantly higher than for the high vagueness group (23\% vs. 12\%, p = .007) . This means that product descriptions with low vagueness can be found significantly better in titles and descriptions. Adding user reviews has two implications: (1) Highly vague product descriptions are better found/covered (12\% vs. 30\%, p < .001) with a significant increase of 18\% in coverage. (2) But also, lowly vague product descriptions are significantly better found (23\% vs. 37\%, p < .001). Product attributes with high and low vagueness are both found in user reviews, resulting in no more statistical difference between the low and high vagueness group.

\subsection{Facet Matching}


Table \ref{tab:results_coverages_facets} provides the average matching coverage of facets both from \emph{Amazon} as a general retailer and those of the specialised retailers. \emph{Amazon} provides 27 facets for laptop search and 12 facets for jacket search, while the specialised retailers offer 84 facets for laptop search (\emph{skinflint}) and 14 for jacket search (\emph{next}). As already noted in Table \ref{tab:results_means}, the descriptions of jackets contain more attributes than those of laptops. Not all attributes mentioned in the user-generated descriptions could be matched to available facets, yet, for both domains, more attributes were matched to the facets of a specialised retailer (33\% coverage) compared to the general retailer (18\% coverage). Additionally, only a third of the attributes could be matched to a facet and a facet value of the specialised retailers, with the percentage being even lower for the general retailer.

In the jacket domain, 8 out of 12 facets on \emph{Amazon} focused on the size of the jackets, leaving only 4 facets for filtering other jacket attributes. The facet that was most often matched was the facet describing the colour of the jacket. The specialised retailer, however, had no redundancy in the facets, with the ``Design Feature'' facet being most often matched. Characteristics such as ``quilted'', ``hooded'' or ``padded' as well as style types such as ``Biker'' or ``Parka'' were possible values of this facet. 

In the laptop domain, the brand plays the most important role as a facet. Other popular facets were ``RAM'', ``Screen Size'', and ''Hard Drive Size''. However, although the coverage was higher for facets of the specialised retailer, it was often not possible to select facet values. While the facets contain precise values, (\textit{``8GB''}, \textit{``15"''}, \textit{``Apple''}), participants described more vague ranges: \textit{``small''}, \textit{``reasonable screen size''}, \textit{``min 8GB''}.

Some descriptions contained very vague language, with no attribute being successfully matched to any facet (neither at the general retailer, nor at the specialised retailer):
\begin{quote}
    \textit{Jacket that is warm and comfortable, yet fashionable and will go with most outfits.}
\end{quote}
and:
\begin{quote}
    \textit{A modern up to date laptop with the software that I use on a daily basis}
\end{quote}
In those cases, attributes could not be matched because either the respective facet was missing (as with \textit{``warm''}) or because it was impossible to determine which attribute of the product would bring about the desired characteristic (e.g. \textit{``comfortable''}, \textit{``fashionable''}).

\begin{table}[tb]
	\centering
	\begin{tabular}{l|rrr|r}
	    & \textbf{All} & \textbf{Laptop} & \textbf{Jacket} & \textbf{Domains} \\
	    & \textbf{mean} & \textbf{mean} & \textbf{mean} & \textbf{p-value} \\ \midrule
        facets \textit{Amazon}		& 18\% & 26\% & 10\% & ** .004 \\
        facets \textit{special ret.}		& 33\% & 33\% & 32\% & .208 \\
	    \bottomrule
	\end{tabular}
	\caption{Average coverages for facet lists, with coverage being the percentage of attributes matched per user-generated description. Right column: p-values of test for significance between domains, where * denotes significance at 95\% CI and ** significance at a 99\% CI.}
    \label{tab:results_coverages_facets}
\end{table}

\begin{table}[tb]
	\centering
	\begin{tabular}{lr|lr}
	    \multicolumn{2}{c|}{\textbf{Laptop}} & \multicolumn{2}{c}{\textbf{Jacket}} \\ 
	    \textbf{attribute} & \textbf{count} & \textbf{attribute} & \textbf{count} \\ 
	    \midrule
	     \textbf{purpose} & 4 & \textbf{purpose} & 34 \\
	      &  & \hspace{3pt}warmth / winter & 30 \\
	      &  & \hspace{3pt}other & 4 \\
	     \textbf{appearance} & 17 & \textbf{appearance} & 15 \\
	     \hspace{3pt}design & 6 & \hspace{3pt}cut & 3 \\
	     \hspace{3pt}size & 5 & \hspace{3pt}style & 12 \\
	     \hspace{3pt}portability & 6 &  &  \\
	     \textbf{experiences} & 6 & \textbf{experiences} & 15 \\
	     \hspace{3pt}life time & 3 & \hspace{3pt}life time  & 4 \\
	     \hspace{3pt}brand reputation & 3 & \hspace{3pt}brandedness & 2 \\
	      &  & \hspace{3pt}quality & 3 \\
	      &  & \hspace{3pt}comfort & 6 \\
	     \textbf{software} & 9 & \textbf{accessories} & 7 \\
	      &  & \hspace{3pt}collar charact. & 2 \\
	      &  & \hspace{3pt}hood charact. & 2 \\
	      &  & \hspace{3pt}zipper charact. & 3 \\
	     \textbf{behaviour} & 29 & \textbf{material} & 19 \\
	     \hspace{3pt}up-to-dateness & 7 & \hspace{3pt}outer material & 2 \\
	     \hspace{3pt}ease-of-use & 4 & \hspace{3pt}inner material & 6 \\
	     \hspace{3pt}speed & 7 & \hspace{3pt}weight & 7 \\
	     \hspace{3pt}computation power & 7 & \hspace{3pt}thickness & 4 \\
	     \hspace{3pt}graphics & 2 &  &  \\
	     \hspace{3pt}audio & 2 &  &  \\
	     \textbf{model} & 16 &  &  \\
	    \bottomrule
	\end{tabular}
	\caption{Proposed facet categories for unmatched attributes}
	\label{tab:results_facets_clustering}
\end{table}

\subsection{Derived Facet Suggestions}
Using the attributes that could not be matched to the facets of the general retailer nor to those of the specialised shop, we grouped similar attributes and identified six new attributes for laptops (containing 86\% of the unmatched attributes) and five for jackets accounting for 94\% of the unmatched jacket attributes (see Table \ref{tab:results_facets_clustering}). Three proposed facets could be helpful for both the laptop and the jacket domain: ``purpose'' (e.g. using the laptop for image editing, or a jacket for the winter season), ``appearance'' (e.g. laptop that fits in a bag, whether a jacket is ``fashionable''), and ``experiences'' relating to attributes that need repeated interaction to judge (e.g. battery life, longevity of the jacket's seams, overall quality of the product).

\section{Discussion and Design Implications}
We investigated in a first step how users formulate their information needs in product search (\textbf{RQ 1}). We found a broad range of vagueness in the user description, from quite precise formulations (\textit{``mustard coloured padded jacket''}) to very vague formulations (\textit{``reasonably priced, with a good sized hard drive''}). Although on average, descriptions received a medium vagueness score (M = 4.79), scores are distributed non-normally: either centered around a low vagueness score or centered around a high vagueness score (\textbf{RQ 1.1}). Designers of an interactive system dealing with vague queries should keep this in mind and \textit{provide reliable search results and functions for both cases: very precise queries and highly vague queries.}

Additionally, we examined the influence of the product domain (\textbf{RQ 1.2}), finding that there are significant differences between the laptop and the jacket domain. Therefore, the results probably do not generalise across other domains. There are significantly more attributes mentioned in jacket descriptions than for laptops, while laptop descriptions are slightly more vague. Users therefore need different support in different domains. As there cannot be a one-fits-all system, it is inevitable for the design of an interactive search system to \textit{include the users from an early stage}. An important finding is the weakly negative correlation between satisfaction and amount of attributes mentioned -- the more attributes are mentioned, the lower the satisfaction. Users with a precise conception of their information need were not simultaneously better at finding what they were looking for, despite mentioning more desired attributes. Furthermore, there were differences between the retailer-generated description and the query. Participants used \textit{generalisations, word omissions, and references (pronouns instead of nouns) that have to be resolved when designing for natural language queries}.

The second aim of this study was to determine to what extent available information sources match natural language descriptions of information needs (\textbf{RQ 2}). Especially in voice applications, systems need to be able to process natural language and live up to the user's expectations of a human-human-like conversation. Using the information on the product pages of the selected products, we found that on average, only a fifth of the desired attributes mentioned in the user-generated descriptions were covered by retailer-generated content (i.e., the product page title and the product description). Reviews significantly increase this matching percentage: Adding reviews to the retailer-generated content yields significantly better coverage of desired attributes in both domains. Reviews especially help when matching vague descriptions. They match equally well to low vagueness as they match to high vagueness descriptions, which is not the case for retailer-generated content. Retailer-generated content is significantly worse at matching to highly vague descriptions as compared to low-vagueness descriptions. We suspect that retailers attempt to describe their products precisely to not give a false image of their product, while user reviews are written in natural language with an equal level of vagueness as the user-generated descriptions. Therefore, we suggest to not only rely on retailer-generated content in the retrieval process but to also \textit{include user-generated reviews to handle vague search intents when designing new interactive search systems}. Not only matching algorithms could profit from user-generated content: Shopping assistants and conversational agents in the context of online shopping would be able to process vague search intents.

Current online product search systems also offer facets for filtering, raising the question whether current facets fit to natural language queries (\textbf{RQ 3}). The facets of retailers specialised in the respective product domain match better to the user-generated descriptions than the facets of a general retailer, yet do not cover all attributes mentioned in the user descriptions. We therefore propose to \textit{add more user-centered facets that relate not only to hard facts (like the storage size or the brand), but also to experienced attributes} such as quality or reputation of the brand (see Table \ref{tab:results_facets_clustering}). A fair amount of those suggested facets are difficult to quantify, e.g. quality or longevity, but are of importance to the user. Giving the user the possibility to indicate how important the respective attribute is, as done in previous research with sliders \cite{kern2018evaluation,sciascio2016rank}, could be a way to process those vague requirements. However, selecting the value of the facet is the next hurdle. Participants often described ranges of values (\textit{``at least''}) or more abstract concepts of values, e.g. \textit{``large storage''}. The specialised notebook retailer \emph{skinflint} provides an overwhelming amount of 84 facets with technical attributes. Selecting the correct facet and facet values, while simultaneously keeping the overview over the result list adds cognitive burden on the user. Here, adjusting the ordering of facets according to the user's input could help to reduce the cognitive burden. Besides compiling user-driven facets, \textit{designers of interactive search systems should consider mapping vague facet values onto dynamic ranges}, e.g. assigning \textit{``cheap''} laptop to the lower third of the current price range, or \textit{``fast''} processor onto the upper quartile of available processor speeds. Lexical ambiguity could be processed with query expansion based on synonyms from thesauri and user reviews, all while giving users insights into the system's processing steps and the possibility to correct faulty interpretations.

We approach the topic of information need formulation from the user side and provide empirical evidence showing that search systems could be improved through utilising user-generated content. Systems supporting the user in online shopping or voice search could profit from user-generated content to improve their processing of vague language. For product search facets, we suggest some new facets based on the user needs mentioned in the study. In our experiment, user reviews have shown to bring a substantial improvement for finding information sought in natural language information needs. User-generated content is therefore a valuable source to extract new facets or determine the relevance of a specific product for the user, i.e. by using the reviews during the automatic retrieval and ranking process. The next step, generating facets and facet values from user-generated content has already been investigated \cite{feuerbach2017enhancing,hirschmeier2018social}. Our findings validate the applicability of those approaches and highlight the inaptitude of current search systems to deal with vague natural language. To create modern systems that live up to the expectations of users, search systems need to be designed from a user perspective, supporting natural language and vagueness where appropriate.


\subsection{Limitations}
In this section, we summarize the limitations our user study face. The descriptions of jackets are highly influenced by the season in which the user study took place. As the study was conducted in November, most participants described a winter jacket, limiting the mentioned attributes to this category. Future research should repeat the experiment during a different season to yield a fuller image of desired attributes. Furthermore, 
other product domains should be investigated in further studies.

Our results could furthermore be influenced by the amount of available reviews of a product. While some products had a great amount of reviews (500 reviews for one product), others had little to no reviews (14 products without reviews, others with less than 10 reviews). This limitation also extends to proposed solutions: Using user reviews to improve product search and process vagueness in natural language queries requires the existence of user reviews in the first place.

The analysis of matching percentages (``coverage'') is limited by the choice of purely lexical matching. Compared to manual matching, where annotators use a more sophisticated approach answering ``Is this information available in the text?'' rather than ``Is this string a sub-string of the text?'', the automatic matching delivers conservative results. A more sophisticated matching would account for synonyms, homonyms, generalisations, and ambiguity. To develop a sophisticated algorithm, a deep understanding of vagueness in each product category is needed. Technical terms and technical abbreviations often do not appear in standard synonym databases. As described by Lehtola, Heinecke and Bounsaythip \cite{lehtola2003intelligent}, retailers and users use different vocabularies, likewise complicating the application of advanced natural language processing techniques used in conversational search, e.g. query expansion and the usage of word embeddings \cite{andolina2018investigating}. Various natural language processing methods have already been developed for learning distributed word representations \cite{mikolov2013distributed, pennington2014glove} to address the vocabulary mismatch problem. Word embeddings can for example be used to identify which words are used in the same context. The context, however, is lost when using bullet point lists. For retailer-generated content, which is often a list of technical specifications, word embeddings might not be enough. Contrarily, for user-generated content, word embeddings could be used to identify synonyms and expand the natural language queries with additional, related terms. The embeddings would have to be trained per topic to account for domain-specific vocabulary and word usage \cite{diaz2016query}. Other methods make use of deep neural networks to improve the matching process between queries and products \cite{shen2014learning}. These methods, however, are often based on behavioural data such as click-through data. To develop a search system with deep networks, data about user behaviour would need to be collected and annotated. The potential to apply natural language techniques in the product search scenario needs to be investigated in future work. 

Overall, understanding how users formulate their search intent in natural language provides the basis to develop more sophisticated matching algorithms. Our research provides first insights for the jacket and laptop domain and clears the way for developing sophisticated product search engines.

\section{Conclusion}
This paper investigated intuitively formulated information needs in product search with respect to vagueness and fitness to currently existing product search systems through an online user study (N=132). Our findings show the broad variety of information need formulations and how vagueness is used to describe products. We found that retailer-generated content does not deal well with natural language queries. User reviews have shown to be a valuable source for improving product matching especially for highly vague search queries. User reviews also provide a basis to generate user-centered facets or expand queries with synonymous terms. Currently, based on our findings and the derived design implications, we develop and evaluate a prototype of a search systems that supports vague information needs. In addition to studying the user experience of such a system, we will investigate how to include products without a sufficient amount of user reviews in a system that relies primarily on user-generated content. In this context, we also explore the potential of using more sophisticated natural language processing techniques to improve the matching process.

\section{Acknowledgments}
This work was partly funded by the DFG, grant no. 388815326; the VACOS project at GESIS.

%
%
%
%
%
%
%

\balance{}

\bibliographystyle{SIGCHI-Reference-Format}
\bibliography{sample}

\end{document}